\documentclass[journal,10pt,twocolumn,twoside]{IEEEtran}

\usepackage{amsmath}
\usepackage{nccmath}
\usepackage[utf8]{inputenc} %unicode support
\usepackage{epsfig,scalefnt,multirow}
\usepackage{url}
\usepackage{ifthen}
\usepackage{mathtools}
\usepackage{cite}
\usepackage{graphicx}
\usepackage{amssymb}
\usepackage{tabularx}
\usepackage{amsmath}
\usepackage{epstopdf}
\usepackage{algorithm}
\usepackage{algpseudocode}
\usepackage{algpascal}
\usepackage{cases}
\usepackage{stfloats}
\usepackage{xcolor}
\usepackage{tabularx}% http://ctan.org/pkg/tabularx
\usepackage{soul}
\usepackage{color}
\sethlcolor{green}

  \def\cC{{\mathcal{C}}} 
 \def\cF{{\mathcal{F}}} \def\cG{{\mathcal{G}}} 
   
 \def\cN{{\mathcal{N}}}  
  \def\cS{{\mathcal{S}}} 
  \def\cW{{\mathcal{W}}}

\def\argmax{\mathop{\mathrm{argmax}}}

\def\b0{{\pmb{0}}} 

\def\ba{{\mathbf{a}}}  \def\bc{{\mathbf{c}}} 
 \def\bff{{\mathbf{f}}} \def\bg{{\mathbf{g}}} \def\bh{{\mathbf{h}}}

  \def\bw{{\mathbf{w}}} \def\bx{{\mathbf{x}}}
\def\by{{\mathbf{y}}}   

\def\bA{{\mathbf{A}}}

\def\bU{{\mathbf{U}}}

\DeclarePairedDelimiter\norm{\lVert}{\rVert}

\begin{document}
\title{Polar-Cap Codebook Design for MISO Rician Fading Channels with Limited Feedback}

\author{Sung~Hyuck~Hong,~\IEEEmembership{Student Member,~IEEE,}
	Sucheol~Kim,~\IEEEmembership{Student Member,~IEEE,} \\
	Junil~Choi,~\IEEEmembership{Senior Member,~IEEE,}
	and~Wan~Choi,~\IEEEmembership{Fellow,~IEEE}%<-this % stops a space
%\thanks{Manuscript received April 19, 2005; revised August 26, 2015. \textit{(Corresponding author: Junil Choi.)}}% <-this % stops a space
\thanks{This work was supported in part by the Ministry of Science and ICT (MSIT) of the Korea government under the 
Information Technology Research Center (ITRC) support program (IITP-2020-0-01787) supervised by the Institute of Information \& Communications Technology Planning \& Evaluation (IITP), and in part by the National Research Foundation (NRF) grant funded by the MSIT of the Korea government (No. 2019R1C1C1003638).}
\thanks{S. Hong, S. Kim, and J. Choi are with the School of Electrical Engineering, Korea Advanced Institute of Science and Technology, Daejeon 34141, South Korea (e-mail: shong16@kaist.ac.kr; loehcusmik@kaist.ac.kr; junil@kaist.ac.kr).}
\thanks{W. Choi is with the Department of Electrical and Computer Engineering, Seoul National University (SNU), Seoul 08826, South Korea (e-mail: wanchoi@snu.ac.kr).}}

\maketitle

\begin{abstract}
Most of the prior works on designing codebooks for limited feedback systems have not considered the presence of strong line-of-sight (LOS) channel component. This paper proposes the design of polar-cap codebook (PCC) for multiple-input single-output (MISO) limited feedback systems subject to Rician fading channels. The codewords of the designed PCC are adaptively constructed according to the instantaneous strength of the LOS channel component. Simulation results show that the codebook can significantly enhance the performance of transmit beamforming in terms of received signal-to-noise ratio (SNR).
\end{abstract}


% \fi
\IEEEpeerreviewmaketitle

\section{Introduction}
\IEEEPARstart{T}{ransmit} beamforming is a technique of using multiple antennas to increase the spectral efficiency and received signal-to-noise ratio (SNR) of communication systems, thereby enabling higher data rates and more reliable information transmissions \cite{MIETZNER09}. However, channel state information (CSI) must be available at the transmitter (TX) to reap the full benefits of such technique. While the receiver (RX) can usually estimate the channel from reference signals, acquiring CSI at the TX is often challenging in frequency-division duplexing (FDD) systems where channel reciprocity is absent {\cite{MARZETTA06,MUKKAVILLI03}}. One strategy for overcoming this challenge is to utilize a low-rate feedback channel. If the TX and RX share a set of candidate beamforming vectors, i.e., a codebook, then the RX can decide which beamforming vector is best suited to the estimated channel and send the corresponding index to the TX through the feedback channel \cite{XIA06,LOVE08,LOVE04}.

The fifth generation (5G) communication systems are employing millimeter-wave (mmWave) frequency spectrum in order to meet the ever-increasing demand for higher data throughput \cite{LI17,ALKHATEEB14}. Though the TX can, in theory, obtain accurate CSI by leveraging channel reciprocity in time-division duplexing (TDD) mmWave systems, various issues that arise in practice, e.g., calibration error between transmit/receive radio frequency chains, do not guarantee such successful CSI acquisition \cite{XIE17}. While FDD mmWave systems are more suitable for meeting tight latency constraints and providing backward compatibility, explicit channel estimation or use of feedback channel is necessary to achieve satisfactory performance  \cite{AHMADI19, SHEN16,LEE15}.

While there have been many studies on designing codebooks for limited feedback systems, the majority of them assume channels to be Rayleigh faded \cite{LOVE03,YEUNG05,LOVE06}. However, signals received over the line-of-sight (LOS) path are stronger than those received over non-line-of-sight (NLOS) paths in numerous communication settings, where the performance of a codebook designed under the assumption of Rayleigh fading channels would inevitably be degraded. Such performance loss would become even more severe in upcoming mmWave communication systems as signals are subject to greater diffraction and scattering losses at higher frequencies \cite{HEATH16,RAPPAPORT17}. 

To address such breakdown of Rayleigh fading channel models, codebook designs for Rician fading channels were proposed in \cite{KIM09} and \cite{ZHU10}. However, the codebooks in both works are independent of the instantaneous LOS channel component, which, when considered, can yield significant performance improvement. Up to the authors' knowledge, no prior work has proposed a codebook design that takes into account the instantaneous LOS component of Rician fading channels. Such codebook design will be especially critical in mmWave communication systems, where the role of LOS communication link is further emphasized \cite{RAPPAPORT17}.
    
In this paper, we propose the design of polar-cap codebook (PCC) for multiple-input single-output (MISO) limited feedback systems under Rician fading channels. The PCC is a variation of the transformation-based or polar-cap differential codebook that was adopted in the IEEE 802.16m standard \cite{IEEE01,IEEE02,IEEE03}. By exploiting the CSI available at the RX, the proposed design constructs the codebook that is well suited to the instantaneous channel realization. The updated codebook is made available at the TX via marginal increase of feedback information in addition to the best codeword index.

The remainder of this paper is organized as follows. A system model and a Rician fading channel model are introduced in Section \ref{sec:SystemModel}, and the properties of PCC are explained in Section \ref{sec:PCC}. The design and analysis of the proposed PCC are described in Section \ref{sec:Real-Chordal PCC}. Simulation results are presented in Section \ref{sec:Simulation}, which is followed by conclusions in Section \ref{sec:Conlusions}. 
 
\textbf{Notation:} Vectors and matrices are represented by lower and upper boldface letters. The transpose and conjugate transpose of the matrix $\bA$ are denoted by $\bA^{\mathrm{T}}$ and $\bA^{\mathrm{H}}$. The set of all $m \times n$ complex matrices is represented by ${\mathbb{C}}^{m \times n}$. The expectation operator is written as $\mathbb{E}[\cdot]$, while $|{\cdot}|$ and $\norm{\cdot}_2$ each denote the absolute value of a scalar and the $\ell_2$-norm of a vector. The complex normal distribution with mean $m$ and variance $\sigma^2$ is represented by $\cC \cN(m,\sigma^2)$.

\section{System Model} \label{sec:SystemModel}
This paper considers a MISO limited feedback system where a TX equipped with $N_\text{t}$ antennas communicates with a single-antenna RX. The TX and RX share a common codebook $\cF=\{\bff_1,\bff_2,\ldots,\bff_{2^B}\}$, where $\bff_j \in \mathbb{C}^{N_\text{t}\times1},\norm{\bff_j}_2=1,\forall j \in \{ 1,2,\ldots,2^B \}$, and the number of feedback bits $B$ can be any positive integer. By exploiting reference signals that the TX transmits, the RX estimates the channel $\bh \in {\mathbb{C}^{N_\text{t} \times 1}}$ between the TX and RX.

After estimating $\bh$, the RX decides the best codeword index ${j^*}=\argmax_{j \in \{ 1,2,\ldots,2^B\}}|\bh^{\mathrm{H}}\bff_j|$ and conveys it to the TX through the low-rate feedback channel. Subsequently, the TX uses $\bff_{j^*}$ as the beamforming vector to send a transmit symbol $s$ to the RX. The received signal $y$ at the RX is 
\begin{align}
y=\bh^{\mathrm{H}}\bff_{j^*}s+n,\label{received_signal}
\end{align}   
where $n\sim\cC\cN(0,N_0)$ is additive white Gaussian noise. The received SNR $\rho$ is then given by 
\begin{align}
\rho=\frac{P_s|\bh^\mathrm{H}\bff_{j^*}|^2}{N_0},\label{received_SNR}
\end{align} 
where $P_s=\mathbb{E}[|s|^2]$ denotes the average power per transmit symbol. To accurately compare the effectiveness of different codebook designs, we assume that the RX can perfectly estimate $\bh$ and set the feedback channel to be free of error and delay \cite{MUKKAVILLI03,XIA06}. Also, while all the results developed in this paper can be generalized to arbitrary antenna configurations, we assume for simplicity that $N_\text{t}$ antennas at the TX are arranged as a uniform linear array (ULA).

The channel $\bh$ is modeled by
\begin{align}
\bh&=\sqrt{\frac{K}{K+1}}G_{\text{LOS}}\ba(\theta)+\frac{1}{\sqrt{K+1}}\bh_{\text{NLOS}},\label{channel_model}
\end{align}  
where $K$ is the Rician K-factor and $G_{\text{LOS}} \sim \cC\cN(0,1)$ is the complex small-scale fading gain on the LOS path from the TX to the RX. The array response vector $\ba(\theta)=[1,e^{-j\frac{2\pi d \cos(\theta)}{\lambda_\text{c}}},\ldots,e^{-j\frac{2\pi(N_\text{t}-1)d \cos(\theta)}{\lambda_\text{c}}}]^{\mathrm{T}} \in {\mathbb{C}^{N_\text{t} \times 1}}$ is parameterized by the LOS angle $\theta \in [0,\pi]$ between the LOS path and the antenna array at the TX, where $d$ is the distance between two adjacent antennas in the TX, and $\lambda_\text{c}$ is the wavelength of the carrier signal. Lastly, the NLOS channel vector $\bh_{\text{NLOS}} \in {\mathbb{C}^{N_\text{t} \times 1}}$  is modeled by a vector whose entries are independently and identically distributed with $\cC\cN(0,1)$ \cite{KIM09,ZHU10}.

We assume that the TX and RX can acquire $\theta$, which is a long-term statistic in general, by using various position or angle-of-arrival/angle-of-departure (AoA/AoD) estimation techniques \cite{GEZICI08,ZHANG12}. We also assume that the RX knows the antenna spacing $d$ at the TX. Then, since both the TX and RX are aware of $N_\text{t}$ and $\lambda_\text{c}$, they can reconstruct $\ba(\theta)$.

% or


\section{Properties of Polar-Cap Codebook} \label{sec:PCC}
In this section, we describe the important properties of PCC, which is a slightly modified version of the polar-cap differential codebook in the IEEE 802.16m standard \cite{IEEE01,IEEE02,IEEE03}. The codewords of a PCC are categorized into two types: basis and non-basis. To design a PCC $\cW$ of cardinality $L$, one should first set the basis codeword $\bw_1 \in \mathbb{C}^{N_\text{t} \times 1}$ and radius $\delta$ of $\cW$, where $\bw_1$ satisfies the unit $\ell_2$-norm constraint, i.e., $\norm{\bw_1}_2=1$, and $\delta$ is a positive real constant. Because there is only one basis codeword per PCC, the remaining $L-1$ codewords of $\cW$ are non-basis codewords, which also satisfy the unit $\ell_2$-norm constraint. The PCC $\cW$ is then designed according to the following criteria \cite{CHOI12}.
\begin{enumerate}
	\item The distance between the basis codeword and any non-basis codeword is equal to the radius $\delta$.
	\item The minimum distance between any two non-basis codewords is maximized.
\end{enumerate}
\noindent
In other words, designing $\cW$ is equivalent to solving the optimization problem
\begin{align}
\cW= \argmax_{\cC \in \mathbb{W}(\bw_1,\delta)}\bigg\{\min_{k,l \in \{2,3,\ldots,L\},\medspace k<l}d(\bc_k,\bc_l)\bigg\},\label{PCC_criterion}
\end{align} 
where $\mathbb{W}(\bw_1,\delta)=\{\{\bw_1,\bc_2,\ldots,\bc_L\} \medspace\big\rvert\medspace d(\bw_1,\bc_j)=\delta, \bc_j \in {\mathbb{C}^{N_\text{t} \times 1}},\norm{\bc_j}_2=1, \forall j \in \{2,3,\ldots,L\} \}$ is the set of all possible codebooks consisting of $\bw_1$ and $L-1$ $N_\text{t}$-dimensional vectors whose distance from $\bw_1$ is $\delta$. $d(\bx,\by)$ is a distance measure defined with respect to two $N_\text{t}$-dimensional vectors $\bx$ and $\by$ with unit $\ell_2$-norm. In this paper, the chordal distance is used as the distance measure, i.e., $d(\bx,\by)=\sqrt{1-|\bx^\mathrm{H}\by|^2}$.

In general, the complexity of designing a PCC relies heavily on the distance measure $d(\bx,\by)$. Fortunately, when $d(\bx,\by)=\sqrt{1-|\bx^\mathrm{H}\by|^2}$, the PCC $\cW$ that has $\bw_1$ as its basis codeword and $\delta$ as its radius can be efficiently generated as \cite{CHOI12}
\begin{align}
\cW=\begin{Bmatrix}
\bw_1,
\bU_{\bw_1}\begin{bmatrix} \sqrt{1-\delta^2}\\ \delta \bg_1\\  \end{bmatrix},\ldots,
\bU_{\bw_1}\begin{bmatrix} \sqrt{1-\delta^2}\\ \delta \bg_{L-1}\\ \end{bmatrix}
\end{Bmatrix}.\label{simple_design_PCC}
\end{align}
Here, $\bU_{\bw_1} \in \mathbb{C}^{N_\text{t} \times N_\text{t}}$ represents a unitary matrix whose first column is $\bw_1$, and $\cG=\{\bg_1,\bg_2,\ldots,\bg_{L-1}\}$ is a Grassmannian codebook \cite{LOVE03} of cardinality $L-1$, where $\bg_j \in \mathbb{C}^{(N_\text{t}-1)\times1}, \norm{\bg_j}_2=1, \forall j \in \{ 1,2,\ldots,L-1\}$.

\section{Proposed Polar-Cap Codebook} \label{sec:Real-Chordal PCC}

\subsection{Design of Proposed Polar-Cap Codebook}
\begin{figure}[t]
	\centering
	\includegraphics[width=1\columnwidth]{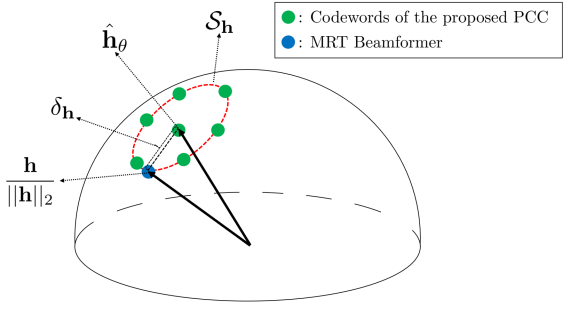}
	%   % where an .eps filename suffix will be assumed under latex,
	%   % and a .pdf suffix will be assumed for pdflatex
	\caption{Visual representation of the proposed PCC.}\label{Real-ChordalPCC_visual}
\end{figure}

In this subsection, we explain the design of the proposed PCC and delineate how it can be implemented in MISO limited feedback systems. The proposed PCC has $\hat{\bh}_\theta$ as its basis codeword and $\delta_\bh$ as its radius, where
\begin{align} \hat{\bh}_\theta&=\frac{\ba(\theta)}{\norm{\ba(\theta)}_2} \notag\\ 
&=\frac{1}{\sqrt{N_\text{t}}}[1,e^{-j\frac{2\pi d \cos(\theta)}{\lambda_\text{c}}},\ldots,e^{-j\frac{2\pi(N_\text{t}-1)d \cos(\theta)}{\lambda_\text{c}}}]^{\mathrm{T}},\label{basis_codeword_RealPCC} \\
\delta_\bh&=d\left(\frac{\bh}{\norm{\bh}_2},\hat{\bh}_\theta\right)=\sqrt{1-\bigg|\frac{\bh^{\mathrm{H}}}{\norm{\bh}_2}\hat{\bh}_\theta \bigg|^2}. \label{radius_RealPCC}
\end{align}
Fig. \ref{Real-ChordalPCC_visual} shows a visual representation of the proposed PCC. The radius $\delta_\bh$ of the codebook is chosen as the chordal distance between the basis codeword $\hat{\bh}_\theta$ and the maximum ratio transmission (MRT) beamformer $\bh/ \norm{\bh}_2$, which is optimal in the sense of maximizing the received SNR. When LOS channel component is dominant, i.e., $\delta_\bh$ is small, the codewords of the PCC are designed such that they are located close to $\hat{\bh}_\theta$ and thus quantize the current channel more effectively. When LOS channel component is not as strong, i.e., $\delta_\bh$ is large, the codebook is constructed with vectors that are less correlated with $\hat{\bh}_\theta$ in order to compensate for the degradation of LOS communication link. If the proposed PCC contains $2^B$ codewords, then its non-basis codewords can be interpreted as a collection of $2^B-1$ vectors in the set $\cS_\bh= \bigg\{\bx \in \mathbb{C}^{N_\text{t} \times 1} \medspace\big\rvert\medspace \norm{\bx}_2=1, {\sqrt{1-|\bx^H\hat{\bh}_\theta|^2}}= \delta_\bh \bigg\}$, which also includes $\bh/ \norm{\bh}_2$ as illustrated in Fig. \ref{Real-ChordalPCC_visual}.

Now we explain how the proposed codebook can be applied to the system model defined in Section \ref{sec:SystemModel}. First, we assume that the TX and RX initially share a Grassmannian codebook  $\cG=\{\bg_1,\bg_2,\ldots,\bg_{2^B-1}\}$, where $\bg_j \in \mathbb{C}^{(N_\text{t}-1)\times1},\norm{\bg_j}_2=1,\forall j \in \{ 1,2,\ldots,2^B-1\}$. Next, using its knowledge of $\ba(\theta)$ and $\bh$, the RX constructs the PCC $\cF(\bh)=\{\bff_1,\bff_2,\ldots,\bff_{2^B}\}$ whose basis codeword is $\hat{\bh}_\theta$ in (\ref{basis_codeword_RealPCC}) and radius is $\delta_\bh$ in (\ref{radius_RealPCC}). From (\ref{simple_design_PCC}), the codebook $\cF(\bh)$ can be expressed as
\begin{align}
\cF(\bh)=\begin{Bmatrix}
\hat{\bh}_\theta,
\bU_{\hat{\bh}_\theta}\begin{bmatrix} \sqrt{1-\delta^2_\bh}\\ \delta_\bh \bg_1\\  \end{bmatrix},\ldots,
\bU_{\hat{\bh}_\theta}\begin{bmatrix} \sqrt{1-\delta^2_\bh}\\ \delta_\bh \bg_{2^B-1}\\ \end{bmatrix}
\end{Bmatrix},
\end{align}
where the notation $\cF(\bh)$ is used to highlight the adaptiveness of the codebook to $\bh$. After constructing $\cF(\bh)$, the RX conveys $\delta_\bh$ and the best codeword index $j^*=\argmax_{j \in \{ 1,2,\ldots,2^B\}}|\bh^{\mathrm{H}}\bff_j|$ to the TX through the feedback channel. Finally, the TX, which is also aware of $\ba(\theta)$, generates the beamforming vector $\bff_{j^*}$, where
\begin{align}
	\bff_{j^*}=
	\begin{cases}
	\hat{\bh}_\theta &\text{if } j^*=1, \\[2ex]
	\bU_{\hat{\bh}_\theta}\begin{bmatrix} \sqrt{1-\delta^2_\bh}\\ \delta_\bh \bg_{j^*-1}\\ \end{bmatrix} &\text{otherwise}.
	\end{cases}
\end{align}
Note that the TX can generate $\bff_{j^*}$ without constructing the entire codebook $\cF(\bh)$.

Under the standard assumption that $2^B \geq N_\text{t}$, the computational complexity of the proposed PCC design is given by $\mathcal{O}(2^BN_\text{t}^2)$, whereas the design proposed in \cite{KIM09} and \cite{ZHU10} each have the complexity of $\mathcal{O}(2^BN_\text{t})$ and $\mathcal{O}(2^BN_\text{t}^2)$. As will be verified in Section \ref{sec:Simulation}, the PCC design achieves significant performance gains over both of them by adaptively adjusting the codewords according to LOS channel component.

\subsection{Analysis of Proposed Polar-Cap Codebook} \label{section:4B}
In this subsection, we investigate the adaptive characteristic of the proposed PCC by deriving an expression that approximates the conditional mean of the radius $\delta_\bh$ of the codebook given a value of the small-scale fading gain $G_{\text{LOS}}$ on the LOS path. The expression gives insight into how the structure of the codebook varies with the channel parameters $K$ and $G_{\text{LOS}}$.

By rewriting (\ref{radius_RealPCC}) as $\delta_\bh=\sqrt{1-\frac{|\bh^{\mathrm{H}}\hat{\bh}_\theta|^2}{\norm{\bh}_2^2}}$, we can view $\delta_\bh$ as a function of two random variables $|\bh^{\mathrm{H}}\hat{\bh}_\theta|^2$ and $\norm{\bh}_2^2$. Given $G_{\text{LOS}}=g$, simple algebra reveals that 
\begin{align}
\mathbb E[|\bh^{\mathrm{H}}\hat{\bh}_\theta|^2 \medspace\rvert\medspace G_{\text{LOS}}=g]&=\frac{KN_\text{t}|g|^2+1}{K+1},\label{Mean_num}\\ 
\mathbb E[\norm{\bh}_2^2 \medspace\rvert\medspace G_{\text{LOS}}=g]&=\frac{KN_\text{t}|g|^2+N_\text{t}}{K+1}, \label{Mean_den}
\end{align}
where $\mathbb E[X \medspace\rvert\medspace G_{\text{LOS}}=g]$ is the conditional mean of a random variable $X$ given $G_{\text{LOS}}=g$.
Further computation shows that 
\begin{align}
CV_{|\bh^{\mathrm{H}}\hat{\bh}_\theta|^2 \medspace\rvert\medspace g} &=\sqrt{\frac{2KN_\text{t}|g|^2+1}{(KN_\text{t}|g|^2+1)^2}},\label{CV_num} \\
CV_{||\bh||_2^2 \medspace\rvert\medspace g}&=\sqrt{\frac{2K|g|^2+1}{N_\text{t}(K|g|^2+1)^2}}, \label{CV_den}
\end{align}
where $CV_{X \medspace\rvert\medspace g}=\frac{\sqrt{\mathbb E[|X-\mathbb E[X \medspace\rvert\medspace G_{\text{LOS}}=g]|^2 \medspace\rvert\medspace G_{\text{LOS}}=g]}}{\mathbb E[X \medspace\rvert\medspace G_{\text{LOS}}=g]}$ denotes the coefficient of variation of a random variable $X$ given $G_{\text{LOS}}=g$. It can be observed from (\ref{CV_num}) and (\ref{CV_den}) that $CV_{|\bh^{\mathrm{H}} \hat{\bh}_\theta|^2 \medspace\rvert\medspace g} \approx 0$ and $CV_{||\bh||_2^2 \medspace\rvert\medspace g} \approx 0$ when $K^2N^2_\text{t}|g|^4 \gg 0$ and $ K^2N_\text{t}|g|^4+(2K|g|^2+1)(N_\text{t}-1) \gg 0$. Therefore, under the condition that $K^2N^2_\text{t}|g|^4$ and $K^2N_\text{t}|g|^4+(2K|g|^2+1)(N_\text{t}-1)$ are sufficiently greater than $0$, the conditional mean $\mathbb E[\delta_\bh \medspace\rvert\medspace G_{\text{LOS}}=g]$ of $\delta_\bh$ given $G_{\text{LOS}}=g$ can be approximated as
\begin{align}
 &\mathbb E[\delta_\bh \medspace\rvert\medspace G_{\text{LOS}}=g] \nonumber \\ 
 &\quad=\int_0^\infty \! \int_0^\infty \! \sqrt{1-\frac{x}{y}}f_{|\bh^{\mathrm{H}}\hat{\bh}_\theta|^2,||\bh||_2^2 \medspace\rvert\medspace G_{\text{LOS}}}(x,y \medspace\rvert\medspace g)\, \mathrm{d}x \, \mathrm{d}y \nonumber\\
 &\quad=\int_0^\infty \! \int_0^\infty \! \sqrt{1-\frac{x}{y}}f_{|\bh^{\mathrm{H}}\hat{\bh}_\theta|^2 \medspace\rvert\medspace G_{\text{LOS}}}(x \medspace\rvert\medspace g) \nonumber\\
 & \qquad \qquad  \qquad \qquad  \qquad \times f_{\norm{\bh}_2^2 \medspace\rvert\medspace |\bh^{\mathrm{H}}\hat{\bh}_\theta|^2, G_{\text{LOS}}}(y \medspace\rvert\medspace x,g)\, \mathrm{d}x \, \mathrm{d}y \nonumber\\
&\quad\overset{\mathrm{(a)}}{\approx}
\int_0^\infty \! \int_0^\infty \! \sqrt{1-\frac{x}{y}} \delta(x-\mathbb E[|\bh^{\mathrm{H}}\hat{\bh}_\theta|^2 \medspace\rvert\medspace G_{\text{LOS}}=g]) \nonumber\\ 
&\qquad \qquad  \qquad  \qquad  \qquad  \times \delta(y-\mathbb E[\norm{\bh}_2^2 \medspace\rvert\medspace G_{\text{LOS}}=g]) \, \mathrm{d}x \, \mathrm{d}y \nonumber\\
&\quad\overset{\mathrm{(b)}}{=}\sqrt{1-\frac{\mathbb E[|\bh^{\mathrm{H}}\hat{\bh}_\theta|^2 \medspace\rvert\medspace G_{\text{LOS}}=g]}{\mathbb E[\norm{\bh}_2^2 \medspace\rvert\medspace G_{\text{LOS}}=g]}} \nonumber \\
&\quad\overset{\mathrm{(c)}}{=}\sqrt{\frac{N_\text{t}-1}{N_\text{t}(K|g|^2+1)}}.\label{mean_chordal_approx}
\end{align}
Here, $f_{|\bh^{\mathrm{H}}\hat{\bh}_\theta|^2,||\bh||_2^2 \medspace\rvert\medspace G_{\text{LOS}}}(x,y \mspace{4mu}\rvert\mspace{4mu} g)$ is the conditional joint probability density function (PDF) of $|\bh^{\mathrm{H}}\hat{\bh}_\theta|^2$ and $\norm{\bh}_2^2$ given $G_{\text{LOS}}=g$, $f_{|\bh^{\mathrm{H}}\hat{\bh}_\theta|^2 \medspace\rvert\medspace G_{\text{LOS}}}(x \mspace{4mu}\rvert\mspace{4mu} g)$ is the conditional PDF of $|\bh^{\mathrm{H}}\hat{\bh}_\theta|^2$ given $G_{\text{LOS}}=g$, $f_{\norm{\bh}_2^2 \medspace\rvert\medspace |\bh^{\mathrm{H}}\hat{\bh}_\theta|^2,G_{\text{LOS}}}(y \medspace\rvert\medspace x,g)$ is the conditional PDF of $\norm{\bh}_2^2$ given $|\bh^{\mathrm{H}}\hat{\bh}_\theta|^2=x$ and $G_{\text{LOS}}=g$, and $\delta(x)$ is the Dirac delta function. In (\ref{mean_chordal_approx}), (a) follows from the fact that a random variable converges to its mean as its coefficient of variation goes to zero, (b) is derived from the definition of Dirac delta function \cite{HARRIS14}, and (c) is obtained by substituting the expressions in (\ref{Mean_num}) and (\ref{Mean_den}).

For a typical MISO Rician fading channel, where $K$ is large enough to at least ensure that the LOS channel component $\sqrt{\frac{K}{K+1}}G_{\text{LOS}}\ba(\theta)$ is not negligible, it is clear that
\begin{align}
&\mspace{25mu}\mathbb{E}[K^2N^2_\text{t}|G_{\text{LOS}}|^4]=2K^2N^2_\text{t} \gg 0,\\[4pt]
\mathbb{E}[&K^2N_\text{t}|G_{\text{LOS}}|^4+(2K|G_{\text{LOS}}|^2+1)(N_\text{t}-1)] \nonumber \\
&\mspace{10mu}=2K^2N_\text{t}+(2K+1)(N_\text{t}-1) \gg 0.
\end{align}
This suggests that the approximation in (\ref{mean_chordal_approx}) can be justified unless $|g|$ is extremely close to 0. 

The expression in (\ref{mean_chordal_approx}) manifests how the proposed PCC is adaptively designed according to the current channel realization. More specifically, the expression shows that $\mathbb E[\delta_\bh \medspace\rvert\medspace G_{\text{LOS}}=g]$ decreases with increasing $K|g|^2$, which indicates the strength of the LOS channel component. This means that the codewords of the proposed PCC tend to be placed near the basis codeword $\hat{\bh}_\theta$ at large values of $K|g|^2$, while they tend to be spaced far apart at small values of $K|g|^2$ to cope with the relatively strong NLOS paths. The proposed codebook design therefore adjusts the codewords according to how strong the LOS channel component is at the moment, making the PCC highly adaptable to the instantaneous condition of LOS communication link.
\begin{figure}[t]
	\centering
	\includegraphics[width=1\columnwidth]{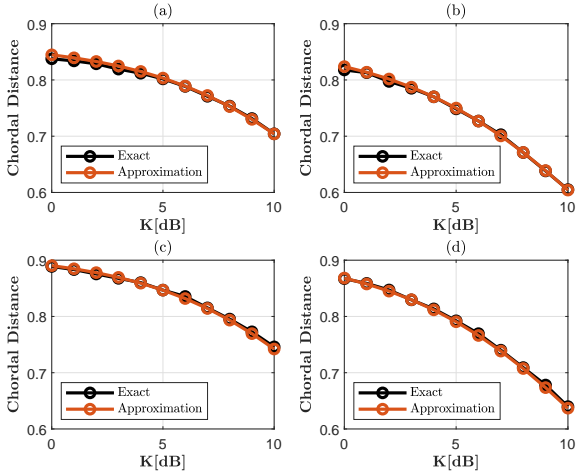}
	%   % where an .eps filename suffix will be assumed under latex,
	%   % and a .pdf suffix will be assumed for pdflatex
	\caption{Comparison of $\mathbb E[\delta_\bh \medspace\rvert\medspace G_{\text{LOS}}=g]$ and its approximation for different values of $K$. (a) $N_\text{t}=4, |G_\text{LOS}|^2=0.0513$. (b) $N_\text{t}=4, |G_\text{LOS}|^2=0.1054$. (c) $N_\text{t}=6, |G_\text{LOS}|^2=0.0513$. (d) $N_\text{t}=6, |G_\text{LOS}|^2=0.1054$.}\label{fig:chordal_approx_nt=4,6}
\end{figure}

\begin{figure}[t]
	\centering
	\includegraphics[width=1\columnwidth]{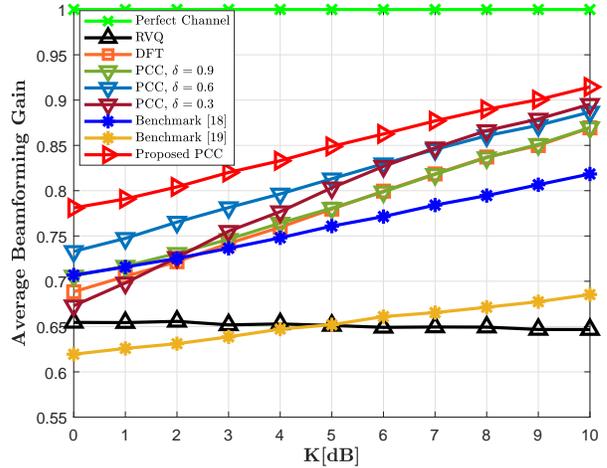}
	%   % where an .eps filename suffix will be assumed under latex,
	%   % and a .pdf suffix will be assumed for pdflatex
	\caption{The average beamforming gain of the proposed PCC and other codebooks as a function of $K$.}\label{fig:BFgain_1st}
\end{figure}

\section{Simulation Results} \label{sec:Simulation}
In this section, we present the results of Monte-Carlo simulations to evaluate and compare the performances of the proposed PCC and other codebooks, which are the random vector quantization (RVQ) codebook \cite{YEUNG05}, discrete Fourier transform (DFT) codebook \cite{WU12}, three fixed-radius PCCs ($\delta=0.9,0.6,0.3$) with $\hat{\bh}_{\theta}$ as the basis codeword, and two benchmark codebooks in \cite{KIM09} and \cite{ZHU10}. The beamforming gain $|\bh^\mathrm{H}\bff_{j^*}|^2/\norm{\bh}^2_2=\max_{j \in \{ 1,2,\ldots,2^B \}}|\bh^{\mathrm{H}}\bff_j|^2/\norm{\bh}^2_2$ is used as the performance metric. The Grassmannian codebooks used to create PCCs and the codebook in \cite{KIM09} are obtained by performing the subspace packing algorithm proposed in \cite{MEDRA14}. We set $N_\text{t}=4, B=4$, and $d=\lambda_\text{c}/2$ for all the simulations.

Fig. \ref{fig:chordal_approx_nt=4,6} compares $\mathbb E[\delta_\bh \mspace{4mu}\rvert\mspace{4mu} G_{\text{LOS}}=g]$ and the derived expression in (\ref{mean_chordal_approx}) for different values of $K$, where the values $0.0513$ and $0.1054$ are the $5$th and $10$th percentile of $|G_{\text{LOS}}|^2$. The figure shows that the expression accurately approximates $\mathbb E[\delta_\bh \mspace{4mu}\rvert\mspace{4mu} G_{\text{LOS}}=g]$, even at small values of $|G_\text{LOS}|$ and $K$. Although not included in this paper, the comparisons with different values of $|G_\text{LOS}|$ and $N_\text{t}$ show similar results.

Fig. \ref{fig:BFgain_1st} shows the average beamforming gain of the codebooks as a function of $K$. The result shows that the proposed PCC outperforms all the other codebooks at each value of $K$, verifying that higher beamforming gain can be achieved by adjusting the radius of PCC according to the strength of the LOS channel component. While the fixed-radius PCC with $\delta=0.3$ performs worse than the other two fixed-radius PCCs at $K$ smaller than 3 dB, it outperforms both of them as $K$ increases. This indicates that, as discussed in Section \ref{section:4B}, a PCC with a small radius tends to perform better when $K$ is sufficiently large.

\begin{figure}[t]
	\centering
	\includegraphics[width=1.1\columnwidth]{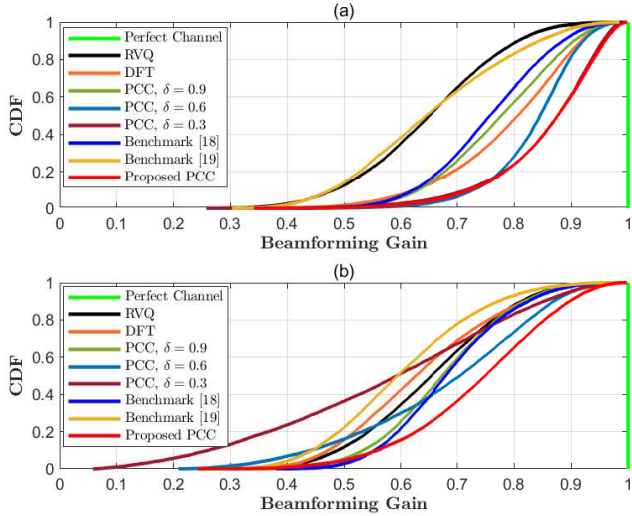}
	%   % where an .eps filename suffix will be assumed under latex,
	%   % and a .pdf suffix will be assumed for pdflatex
	\caption{The cumulative distribution function of the beamforming gain of the proposed PCC and other codebooks with $K=5$ dB. (a) $|G_\text{LOS}|^2=0.6931$. (b) $|G_\text{LOS}|^2=0.1054$.}\label{fig:BFgain_CDF}
\end{figure}

To evaluate the impact that the small-scale fading of the LOS path has on the performance of the proposed PCC, the cumulative distribution function (CDF) of the beamforming gain of the codebook is plotted in Fig. \ref{fig:BFgain_CDF} where $K=5$ dB. Figs. \ref{fig:BFgain_CDF} (a) and (b) respectively show the result when $|G_\text{LOS}|^2=0.6931$ and $|G_\text{LOS}|^2=0.1054$, which are the $50$th and $10$th percentile of $|G_\text{LOS}|^2$. As shown in the figures, the beamforming gain of the proposed PCC has the greatest median at both values of $|G_\text{LOS}|^2$. Note that the fixed-radius PCC with $\delta=0.3$ performs nearly as good as the proposed PCC at $|G_\text{LOS}|^2=0.6931$. However, the codebook is highly vulnerable to small values of $|G_\text{LOS}|^2$, as the median of its beamforming gain is even smaller than that of RVQ codebook at $|G_\text{LOS}|^2=0.1054$. In contrast, the proposed PCC can achieve high beamforming gain at both small and large values of $|G_\text{LOS}|^2$, thereby proving the effectiveness of the proposed codebook design.
\section{Conclusions} \label{sec:Conlusions}
In this paper, we proposed the design of PCC for MISO limited feedback systems subject to Rician fading channels. We explained how the proposed PCC, whose codewords are adaptively adjusted to the instantaneous strength of the LOS channel component, can be implemented in the systems of interest. We also examined the relationship between the proposed PCC and the channel parameters by deriving the expression that accurately approximates the conditional mean of the radius of the codebook. Simulation results proved that the proposed PCC achieves considerably high beamforming gain under Rician fading channels and well withstands the fluctuation of the small-scale fading gain on the LOS path.
\ifCLASSOPTIONcaptionsoff
  \newpage
\fi



%\bibliographystyle{IEEEtran}

\bibliographystyle{IEEEtran}
\bibliography{HONG_WCL2020-1689_reference}


\end{document}